\documentclass[prd,preprint,preprintnumbers,superscriptaddress,nofootinbib]{revtex4-1}
\usepackage[a4paper, hdivide={1.91cm,,1.165cm}, vdivide={1.83cm,,3.6cm}]{geometry}

\usepackage{amstext,amssymb}
\usepackage{amsmath}
\usepackage{graphicx}
\usepackage[hyperfootnotes=true]{hyperref}
\usepackage{xspace}
\usepackage{color}
\usepackage{slashed}

\begin{document}
\title{Neutrino mass and charged lepton flavor violation in an extended left-right symmetric model}
\author{Chayan Majumdar, Supriya Senapati, S. Uma Sankar, Urjit A. Yajnik}
\email{\\chayan@phy.iitb.ac.in \\
supriya@phy.iitb.ac.in\\
uma@phy.iitb.ac.in \\
yajnik@iitb.ac.in}
\affiliation{Department of Physics, Indian Institute of Technology Bombay, Powai, Mumbai, Maharashtra 400 076, India}

\begin{abstract}
We consider an $U(1)_{L_\mu -L_\tau}$ extended left-right symmetric gauge theory where the neutrino masses are generated through inverse seesaw mechanism. In this model the muon $(g-2)$ anomaly is accounted for by the mediation of $Z_{\mu\tau}$, the gauge boson of $U(1)_{L_\mu - L_\tau}$ symmetry. The symmetries of the model require the light neutrino mass matrix to have a particular two-zero texture, which leads to non-trivial constraints in the minimum neutrino mass. In addition, the model predicts observable charged lepton flavor violation in $\mu-\tau$ sector.
\end{abstract}

\maketitle

\section{Introduction}
\label{sec:intro}
The prediction of the anomalous magnetic moment $(g-2)$  is one of the triumphs of Quantum Field Theory (QFT) \cite{Schwinger:1948iu, Gell-Mann:1954wra}. The precise measurement of muon $(g-2)$ \cite{Muong-2:2006rrc} revealed a tiny discrepancy with the standard model (SM) prediction. This deviation indicates potential existence of new physics 
\cite{Ajaib:2014ana,Davoudiasl:2014kua,Rentala:2011mr,Kelso:2013zfa, Ky:2000ku,deSPires:2001dmo,Agrawal:2014ufa}. The muon $(g-2)$ anomaly is one of the most compelling reasons for the search for physics beyond the SM. At present, the discrepancy is at the 4.2$\sigma$ level 
\cite{Muong-2:2021ojo}. Extensive studies, both on experimental \cite{Muong-2:2021ojo,Muong-2:2021xzz,Korostelev:IPAC2016-WEPMW001,Gray:2015qna,Venanzoni:2014ixa,Ishida:2009zz,Iinuma:2011zz,
Saito:2012zz,Eads:2015arb,Sakuma:2021svs} and theoretical \cite{Aubin:2012me,Aubin:2013daa,
Aubin:2013yba,Blum:2013xva,Golterman:2013vna,Jegerlehner:2013sja,Nyffeler:2013spo,
Steinhauser:2013gia} frontiers, are being carried out with the aim of improving the precision 
of both the measured value and the SM prediction of muon $(g-2)$. 

The anomalous magnetic moment is characterized by the quantity $a_\mu = (g-2)/2$. 
The present theoretical prediction of $a_\mu$ from SM is \cite{Aoyama:2020ynm}
\begin{equation}
    a_\mu^{\text{SM}} = 116591810(43) \times 10^{-11}.
\end{equation}
It is in disagreement with the nearly two decade old Brookhaven muon $(g-2)$ collaboration (BNL) 
result \cite{Muong-2:2006rrc}
\begin{equation}
    a_\mu^{\text{BNL}} = 116592089(63) \times 10^{-11},
\end{equation}
with $\Delta a_\mu = (287 \pm 80) \times 10^{-11}$ at $3.7 \sigma$ discrepancy.
The theoretical prediction of $a_\mu$ in the SM is a sum of contributions coming 
from Quantum Electrodynamics (QED), electroweak and hadronic sectors :
\begin{equation}
    a_\mu^{\text{SM}} =  a_\mu^{\text{QED}} +  a_\mu^{\text{electroweak}} +  
a_\mu^{\text{hadronic}}.
\end{equation}
Among these three contributions, the QED and the electroweak contributions have been verified 
to high precision \cite{Gnendiger:2013pva,Aoyama:2017uqe}. Therefore, it is possible that 
the discrepancy arises due some unknown loop contributions to $a_\mu^{\text{hadronic}}$
\cite{Davier:2017zfy,Keshavarzi:2018mgv,Davier:2019can}. Another possibility is that the 
discrepancy is caused by new physics at TeV scale. Recently, the Fermilab muon $(g-2)$ 
collaboration (FNAL) has announced an improved measurement 
\cite{Muong-2:2021ojo}
\begin{equation}
    a_\mu^{\text{FNAL}} =  116592040(54) \times 10^{-11}.
\end{equation}
This new result confirms the BNL measurement and increases the extent of discrepancy to 
$4.2 \sigma$ level with $\Delta a_\mu = (251 \pm 59) \times 10^{-11}$. 
On the theoretical frontier, a proposed experiment MUonE \cite{Abbiendi:2016xup}, 
aims to reduce the theoretical uncertainty in $a_\mu^{\text{hadronic}}$ by directly measuring 
the hadronic vacuum polarization more precisely. The theoretical studies to 
account for the muon $(g-2)$ anomaly can be found in references 
\cite{Jegerlehner:2009ry,Lindner:2016bgg,Ajaib:2014ana,Davoudiasl:2014kua,Rentala:2011mr,
Kelso:2013zfa,Ky:2000ku,deS.Pires:2001da,Agrawal:2014ufa,Endo:2013lva,Ibe:2013oha,Everett:2001tq,
Arnowitt:2003vw,Martin:2002eu,Taibi_2015,Altmannshofer:2016brv,Megias:2017dzd,Jana:2020pxx,
Yamaguchi:2016oqz,Yin:2016shg,Endo:2019bcj,Dev:2017fdz}. 

To address this anomaly most of the recent studies \cite{Garani:2019fpa, Biswas:2016yan, 
Biswas:2016yjr, Poddar:2019wvu, Escudero:2019gzq, Dror:2019uea, Dror:2020fbh,Sirunyan:2018nnz,
Araki:2017wyg} focus on new physics governed by $U(1)_{L_\mu -L_\tau}$ symmetry. While the 
total lepton number, $L$, is a sum of individual lepton numbers $L_e$, $L_\mu$ and $L_\tau$, 
one can always choose the difference between any pair of the lepton numbers, such as $L_e - L_\mu$ or 
$L_\mu -L_\tau$ or $L_e - L_\tau$, and gauge it to obtain an anomaly free theory. 
Of these, known phenomenology rules out any but the gauged $U(1)_{L_\mu -L_\tau}$ symmetry. 
The parameters associated with the new  gauge boson, 
$Z_{\mu \tau}$, are not constrained by lepton and hadron colliders since it does not couple to 
electrons and quarks. Many of the new physics scenarios have explored the $U(1)_{L_\mu -L_\tau}$ 
extension of the SM in the context of neutrino masses and mixing, muon $(g-2)$ anomaly, dark matter
and so on \cite{Garani:2019fpa, Heeck:2011wj, Biswas:2016yan, Biswas:2016yjr, Poddar:2019wvu, 
Escudero:2019gzq, Araki:2014ona, Heeck:2010pg, Dror:2019uea, Dror:2020fbh, Sirunyan:2018nnz,Araki:2017wyg,Baek:2001kca,Gninenko:2001hx, Ma:2001md, Choubey:2004hn,Borah:2020jzi,Borah:2021jzu,Borah:2021khc}. By comparison the 
$U(1)_{L_\mu -L_\tau}$ extended left-right symmetric theories have been less explored. The 
left-right symmetric model (LRSM) \cite{Mohapatra:1974gc,Pati:1974yy,Senjanovic:1975rk,
Senjanovic:1978ev,Mohapatra:1979ia,Mohapatra:1980yp,Pati:1973uk,Pati:1974vw} is one of the 
successful beyond SM scenarios, which gives an unified answer to the origin of small neutrino 
masses as well as parity violation in low-energy weak interactions. LRSM naturally hosts a 
right-handed neutrino and offers rich phenomenological aspects in the context of explaining 
neutrino mass, lepton number violation (LNV), lepton flavor violation (LFV) and so on. 

In this paper, we consider an $U(1)_{L_\mu -L_\tau}$ extended LRSM described in
\cite{Majumdar:2020xws}. We explore the constraints on the small neutrino masses
in this model, arising from the new physics which explains the muon $(g-2)$ anomaly.
In manifest LRSM the neutrino masses are generated through type-I+II seesaw mechanism, 
thus providing a very high scale for the right-handed symmetry breaking ($> 10^{14}$ GeV) 
which is far beyond present collider reach. However, addition of extra particles to LRSM 
allows the generation of neutrino mass at a few TeV scale by low-scale seesaw mechanism 
such as linear seesaw, inverse seesaw, double seesaw and so on 
\cite{Mohapatra:1986bd,Akhmedov:1995vm,Akhmedov:1995ip,Barr:2003nn,Nomura:2019xsb,Sruthilaya:2017mzt,Deppisch:2015cua,Humbert:2015yva,Parida:2012sq,Tello:2010am,Barry:2013xxa,Dev:2013vxa,Nemevsek:2011hz,Dev:2014iva,Das:2012ii,Bertolini:2014sua,Borah:2013lva,Chakrabortty:2012mh,Deppisch:2015cua,Majumdar:2018eqz,Bambhaniya:2015ipg,Dev:2014xea,Ezzat:2021bzs,Ashry:2022maw}. In our previous work, described in \cite{Majumdar:2020xws}, we have considered the LRSM inverse 
seesaw (LISS) scenario for the generation of neutrino masses. The symmetries of this model impose severe constraints on the structure of the light neutrino mass matrix and restrict the allowed values of the lightest neutrino mass and the CP-violating phase $\delta$. In addition, the model also allows charged lepton flavor violation at a level observable in near future.

The rest of the paper is organised as follows. In section \ref{sec:mod} we 
outline the model of the $U(1)_{L_\mu -L_\tau}$ extended LRSM. In section \ref{sec:muonanomaly} we explore which of the contributions of the model to muon $(g-2)$ can explain the observed anomaly. In section \ref{sec:LFV}, we study the decay $\tau \rightarrow \mu \gamma$ and in section \ref{sec:numass} we study the model constraints on light neutrino masses and the CP-violating phase. We present our conclusion in the last section.

\section{The Model}
\label{sec:mod}
The model is an $U(1)_{L_\mu -L_\tau}$ extended left-right symmetric theory with the gauge group 
defined as
\begin{equation}
    \mathbb{G}_{\text{LR}}^{\mu \tau} \equiv U(2)_L\times SU(2)_R\times 
U(1)_{B-L} \times SU(3)_C \times U(1)_{L_{\mu}-{L_\tau}}.
\end{equation}
The particle content of the model is given in table 
\ref{tab:particlecontentELISS}.
\begin{table}[htb]
\begin{center}
\begin{tabular}{|c||c|c|c|c|c|c|}
 \hline
          & Fields      & $ SU(2)_L$ & $SU(2)_R$ & $B-L$ & $SU(3)_C$ & 
$U(1)_{L_\mu-L_\tau}$ \\
\hline
 Fermions &$q_L$     &  2         & 1         & 1/3   & 3 & 0  \\
 & $q_R$     &  1         & 2         & 1/3   & 3 & 0  \\
 & $\ell_{e_L}$  &  2         & 1         & -1    & 1  & 0 \\
 & $\ell_{\mu_L}$  &  2         & 1         & -1    & 1  & 1 \\
 & $\ell_{\tau_L}$  &  2         & 1         & -1    & 1  & -1 \\
 & $\ell_{e_R}$  &  1         & 2         & -1    & 1  & 0 \\
 & $\ell_{\mu_R}$  &  1         & 2         & -1    & 1  & 1 \\
 & $\ell_{\tau_R}$  &  1         & 2        & -1    & 1  & -1 \\ 
 \hline
Extra Steriles & $S_{eL}$   &  1         & 1         &  0     & 1  & 0 \\
& $S_{\mu L}$   &  1         & 1         &  0     & 1  & 1 \\
 & $S_{\tau L}$   &  1         & 1         &  0     & 1  & -1 \\
 \hline \hline
 Scalars & $\Phi$   &  2         & 2         &  0     & 1  & 0 \\
  & $H_L$    &  2         & 1         & 1     & 1   & 0 \\
  & $H_R$    &  1         & 2         & 1     & 1   & 0 \\
  Extra Scalar & $\chi$    &  1        & 1         & 0     & 1  & 1 
  \\
\hline
\end{tabular}
\end{center}
\caption{Particle content of the left-right symmetric theory extended with 
$U(1)_{L_{\mu}-{L_\tau}}$ gauge symmetry. The model contains three set of extra 
sterile fermions ($S_L$) and one extra scalar $(\chi)$ along with the usual fermions 
and scalars present in it.}
\label{tab:particlecontentELISS}
\end{table}

We have considered a doublet-variant LRSM in this work \cite{Majumdar:2020xws}. Apart from the 
usual fermions and scalars present in the model, it contains a set of three extra fermions 
which are sterile and one extra scalar. The extra scalar which is singlet under left-right
symmetry is non-trivially charged under $U(1)_{L_{\mu}-{L_\tau}}$ and helps to break the 
$U(1)_{L_\mu -L_\tau}$ symmetry. The extra sterile fermions help to generate the neutrino masses 
through LISS mechanism. We have termed this scenario as extended LRSM inverse seesaw (ELISS) 
scenario. More details about the choice and advantages of the model can be found in 
\cite{Majumdar:2020xws}. 
 
In the scheme, the non-zero \textit{vev} of $H_R$ breaks the left-right symmetry to SM while 
$H_L$ is required for left-right invariance. Further, the spontaneous symmetry breaking (SSB) 
of SM to low energy theory occurs when the scalar bidoublet $\Phi$ takes a non-zero \textit{vev} 
and that generates masses for charged leptons and quarks.  This is as in the usual LRSM. 
Additionally, the \textit{vev} of $\chi$ accomplishes the SSB of $U(1)_{L_\mu -L_\tau}$. 
The \textit{vev} structure of the Higgs spectrum is as follows :
\begin{equation}
    \langle H_R \rangle = \begin{pmatrix}
0 \\
v_R
\end{pmatrix},~~
\langle H_L \rangle = \begin{pmatrix}
0 \\
v_L
\end{pmatrix},~~
\langle \Phi \rangle = \begin{pmatrix}
v_1 & 0 \\
0 & v_2 e^{i\alpha}
\end{pmatrix},~~ \langle \chi \rangle = u,
\label{eqn:vevs}
\end{equation}
 where $\alpha$ is the relative phase between the two \textit{vev}s of the bidoublet.
For the usual particle content of the double-variant LRSM the allowed Yukawa interactions for the leptons are given by,
\begin{align}
    & -\mathcal{L}_{Yuk} \supset 
  \overline{\ell_{e_L}} \left[Y_\ell \Phi 
+ \tilde{Y}_\ell \widetilde{\Phi} \right] \ell_{e_R}
+ \overline{\ell_{\mu_L}} \left[Y_\ell \Phi 
+ \tilde{Y}_\ell \widetilde{\Phi} \right] \ell_{\mu_R}
+ \overline{\ell_{\tau_L}} \left[Y_\ell \Phi + \tilde{Y}_\ell \widetilde{\Phi} \right] \ell_{\tau_R}
+\mbox{h.c.}
\label{eqn:LYukdoubletLRSM}
\end{align}
with $\tilde{\Phi} = \sigma_2 \Phi^{\ast} \sigma_2$ where $\sigma$'s are the Pauli matrices. 
Then, after SSB of $SU(2)_R$ and $SU(2)_L$,
the masses for charged leptons and the Dirac masses for neutrinos can be expressed as 
\begin{equation}
      M_\ell \simeq \tilde{Y}_\ell v_1, ~~~~~~ M_D^\nu = Y_\ell v_1 + \tilde{Y_\ell} v_2  
      e^{-i \alpha}
      \simeq v_1\left(Y_\ell+M_\ell 
      \frac{v_2  e^{-i \alpha}}{v^2_1} \right).
      \label{eq:MD}
\end{equation}
with $Y_\ell \ll \tilde{Y}_\ell$, $v_2 \ll v_1$. The $U(1)_{L_\mu-L_\tau}$ symmetry of our framework constraints $Y_\ell$ and $\tilde{Y}_\ell$ to be diagonal \cite{Majumdar:2020xws}.

The sterile fermions of the ELISS scenario then provide additional contributions to the neutrino mass 
matrix. With charge assignments as in table \ref{tab:particlecontentELISS} the relevant 
$U(1)_{L_\mu - L_\tau}$ invariant Yukawa interaction Lagrangian in this scheme is given by,
\begin{equation}
     -\mathcal{L}_{\text{ELISS}} \supset \mathcal{L}_{\nu_L N_R}+\mathcal{L}_{N_R S_L}+\mathcal{L}_{S_L S_L} 
+ \mathcal{L}_\chi . 
\end{equation}
The neutrino Dirac mass matrix $M_D^\nu$ results from $\mathcal{L}_{\nu_L N_R}$.

The mixing between the fields $N_R$ and $S_L$ arises from 
\begin{equation}
\mathcal{L}_{N_R S_L} \supset Y_{RS}\overline{\ell} \tilde{H}_R S_L = Y_{RS} \langle \tilde{H}_R \rangle \left[\overline{\ell_{e_R}} S_{e_L}+\overline{\ell_{\mu_R}}S_{\mu_L}+\overline{\ell_{\tau_R}} S_{\tau_L}\right].
\label{eqn:LNS}
\end{equation}
The matrix $Y_{RS} \langle \tilde{H}_R \rangle \equiv M$ is also constrained to be diagonal due to $U(1)_{L_\mu-L_\tau}$ symmetry.

The bare Majorana mass term for extra steriles $S_L$ is given by,
\begin{align}
\mathcal{L}_{S_L S_L} &= S^T_L \mu_S  S_L = \left[\mu_{ee} S_{e_L}^T S_{e_L} +\mu_{\mu \tau} S_{\mu_L}^T S_{\tau_L} + \mu_{\mu\tau} S_{\tau_L}^T S_{\mu_L} \right].
\label{eqn:LSS}
\end{align}
where the bare Majorana mass matrix can be expressed as,
\begin{equation}
\mu_S^{bare} = \begin{pmatrix}
\mu_{ee} & 0 & 0 \\
0 & 0 & \mu_{\mu\tau} \\
0 & \mu_{\mu\tau} & 0
\end{pmatrix}.
\end{equation}
The form of $\mu_S^{bare}$ is dictated by $U(1)_{L_\mu-L_\tau}$ symmetry. The extra scalar $\chi$, needed for the SSB of $U(1)_{L_\mu-L_\tau}$, couples to the sterile fields as
\begin{align}
    \mathcal{L}_{\chi}  \supset   \mu_{e\mu}S^T_{e_L}S_{\mu_L}\chi^{\ast} + \mu_{e\tau}S^T_{e_L}S_{\tau_L}\chi + \mu_{e\mu}S^T_{\mu_L}S_{e_L}\chi^{\ast} + \mu_{e\tau}S^T_{\tau_L}S_{e_L}\chi.
    \label{eqn:LchiS}
\end{align}
When $\chi$ acquires a \textit{vev} the above Lagrangian leads to a matrix $\mu_S^{SSB}$, which is of the form 
\begin{equation}
   \mu_S^{SSB} = \begin{pmatrix}
0 & \mu_{e\mu} & \mu_{e\tau} \\
\mu_{e\mu} & 0 & 0 \\
\mu_{e\tau} & 0 & 0
\end{pmatrix}.  
\end{equation}
The total $\mu_S$ matrix, whose elements are in general complex, is given by
\begin{equation}
\mu_S = \mu_S^{bare} + \mu_S^{SSB} = \begin{pmatrix}
\mu_{ee} & \mu_{e\mu} & \mu_{e\tau} \\
\mu_{e\mu} & 0 & \mu_{\mu\tau} \\
\mu_{e\tau} & \mu_{\mu\tau} & 0
\end{pmatrix}.
\end{equation}

The complete $9 \times 9$ neutral fermion mass matrix in the basis $\{\nu_L,\nu_R^c,S_L\}$ takes the form
\begin{equation}
\mathbb{M} = \begin{pmatrix}
0 & M_D^\nu & 0 \\
M_D^{\nu T} & 0 & M^T \\
0 & M & \mu_S \\
\end{pmatrix}.
\label{whole}
\end{equation}
Thus, for mass hierarchy $M > M_D^\nu \gg \mu_S$, the light neutrinos masses in ELISS scenario is read as as~\cite{Mohapatra:1986aw,Dev:2009aw,Sahu:2004sb,Awasthi:2013ff,Pritimita:2016fgr,Majee:2008mn,Kang:2006sn,Hewett:1988xc,Blanchet:2010kw,Dias:2012xp,Das:2017kkm,Dev:2017fdz},
\begin{equation}
m_{\nu} = M_D^\nu{M}^{-1} \mu_S \left(M_D^\nu{M}^{-1}\right)^T .
\label{eqn:LNMM}
\end{equation}

Parametrizing the diagonal matrices $M_D^\nu$ as dia$(a,b,c)$ and $M$ as dia$(M_{11}, M_{22}, M_{33})$, we find
\begin{equation}
m_{\nu} = \begin{pmatrix}
\frac{a^2 \mu_{ee}}{M_{11}^2} & \frac{ab \mu_{e\mu}}{M_{11}M_{22}} & \frac{ac \mu_{e\tau}}{M_{11}M_{33}} \\
\frac{ab \mu_{e\mu}}{M_{11}M_{22}} & 0 & \frac{bc \mu_{\mu\tau}}{M_{22}M_{33}} \\
\frac{ac \mu_{e\tau}}{M_{11}M_{33}} & \frac{bc \mu_{\mu\tau}}{M_{22}M_{33}} & 0
\end{pmatrix}.
\label{eq:mnu}
\end{equation}

This complex symmetric matrix, when diagonalised by a unitary matrix, gives rise to three non-degenerate eigenvalues. It is possible to have sub-eV scale for these eigenvalues as shown below
\begin{equation}
   \left( \frac{m_\nu}{\mbox{0.1\, eV}}\right) = \left(\frac{M_D^\nu}{\mbox{100\, GeV}} \right)^2 
 \left(\frac{\mu_S}{\mbox{eV}}\right) \left(\frac{M}{100\, \mbox{GeV}} \right)^{-2}\,. 
 \label{eqn:LNMMmassscale}
\end{equation}
Indeed the main advantage we derive from this construction is the sizeable light-heavy neutrino mixing  $( V^{\nu \xi} = M_D^\nu M^{-1}$ $\simeq \mathcal{O} (0.1-1))$ with $M$ at few GeV scale. This plays an important role in the explanation of the muon anomaly as discussed in the next section. For more details of the neutrino mass structure one can refer to \cite{Majumdar:2020xws}.

\section{Model Prediction Muon $(g-2)$ with FNAL data}
\label{sec:muonanomaly}

In the model described in the previous section, the contributions to
$\Delta a_\mu$, arise from the interactions of;
\begin{itemize}
 \item singly charged gauge bosons ($W_L, W_R$) with heavy neutral lepton,
 \item neutral vector boson ($Z_R$) with singly charged leptons,
 \item singly charged scalars with neutral lepton,
 \item neutral scalars with muons,
 \item extra new gauge boson $Z_{\mu\tau}$ with muons.
\end{itemize}

In the following we write down the analytic expressions for each of the contributions and study 
numerically all these contributions to $\Delta a_\mu$. Details of the analytical calculations and 
the numerical comparison with the BNL data are given in \cite{Majumdar:2020xws}.

(a) Contribution due to heavy gauge boson mediation :
\begin{align}
&W_L ~\text{mediation :}~~~ \Delta a_\mu (W_L) \simeq 9.06 \times 10^{-9}~g_L^2  \sum_{i=1,..,6} |V_{\mu i}^{\nu \xi}|^2,
\label{eqn:anaexpWL}\\
&W_R ~\text{mediation :}~~~ \Delta a_\mu (W_R) \simeq 2.3 \times 10^{-11} \left( \frac{g_R}{g_L}\right)^2 \left(  \frac{1~\text{TeV}}{m_{W_R}}\right)^2,
\label{eqn:anaexpWR}\\
&Z_R ~\text{mediation :}~~~ \Delta a_\mu (Z_R) \simeq -\frac{1}{4\pi^2}\frac{m^2_\mu}{m^2_{Z_R}}\left[ \left(-\frac{1}{3}\right)|g_{v}^\mu|^2  + \left(\frac{5}{3}\right)|g_{a}^\mu|^2 \right]
\label{eqn:anaexpZR}.
\end{align}
with $m_{W_L}, m_{W_R}, m_{Z_R} \gg m_\mu$ where $m_{W_L}$, $m_{W_R}$, $m_{Z_R}$ and $m_{\mu}$ are 
the respective masses for $W_L$, $W_R$, $Z_R$ and muon and $g_L$ and $g_R$ are the respective gauge 
couplings of $SU(2)_L$ and $SU(2)_R$. In our previous work \cite{Majumdar:2020xws} it was shown that the contribution from Eq. \ref{eqn:anaexpWL} can explain the muon $g-2$ anomaly by itself provided the light-heavy neutrino mixing factor, $V_{\mu i}^{\nu \xi}$, is in the range
$\mathcal{O}(0.7-1)$. However, such large light-heavy neutrino mixing factors are forbidden by the constraints on the deviation from the unitarity in the mixing of the active flavors \cite{Awasthi:2013ff}. In particular, the constraint on such deviation for muon flavor is very strong $(|V_{\mu i}^{\nu \xi}|^2)/2 \equiv |\eta_{\mu\mu}| \leq 8.0 \times 10^{-4}$ \cite{Antusch:2008tz,Antusch:2009pm,Antusch:2006vwa,Forero:2011pc}. Hence, the contribution of heavy gauge boson exchange to muon $g-2$ is negligibly small ($< 10^{-11}$).

(b) Contribution due to scalar mediation : the scalar sector contributions of this model are coming 
from the Higgs bi-doublet, $\Phi$.
\begin{align}
&\text{Charged Scalar mediation :}~~ \Delta a_\mu (\text{CS}) \simeq 
-\frac{1}{4\pi^2}\frac{m^2_\mu}{m^2_{CS}}\left[|g_{s1}^\mu|^2 
\left(\frac{1}{12} \right) + |g_{p1}^\mu|^2 \left(\frac{1}{12} \right) \right],
\label{eqn:anaexpCS}\\
& \text{Neutral Scalar mediation :} \nonumber \\
&\Delta a_\mu (\text{NS}) \simeq 
\frac{1}{4\pi^2}\frac{m^2_\mu}{m^2_{NS}}\left[|g_{s2}^\mu|^2 
\left(-\frac{7}{12} - \text{log} \left (\frac{m_\mu}{m_{NS}} \right) \right) + |g_{p2}^\mu|^2 
\left(\frac{11}{12}+ \text{log} \left (\frac{m_\mu}{m_{NS}} \right) \right) \right],
\label{eqn:anaexpNS}
\end{align}
where $m_{\text{CS}}$ and $m_{\text{NS}}$ are 
the masses of the charged and the neutral scalars respectively. The couplings $g^\mu_{s1}, g^\mu_{p1}, g^\mu_{s2}$ and $g^\mu_{p2}$ are related to muon Yukawa coupling which is of order $m_\mu / m_{W_L} \sim 10^{-3}$. The lower limits on scalar masses are 
\begin{itemize}
\item $m_{\text{CS}}>$  1.1 TeV from direct search \cite{ATLAS:2018gfm},
\item $m_{\text{NS}}>$  1870 GeV through direct search for CP-even scalars \cite{CMS:2019bnu},
\item $m_{\text{NS}}>$  20 TeV through FCNC for CP-odd scalars \cite{Zhang:2007da}.
\end{itemize} 
Given these small couplings and large masses it is straightforward to show that the maximum possible contribution to $\Delta a_\mu$ from scalar exchange is $\mathcal{O}(10^{-16})$.

\begin{figure}[htb!]
\centering
\includegraphics[scale=0.9]{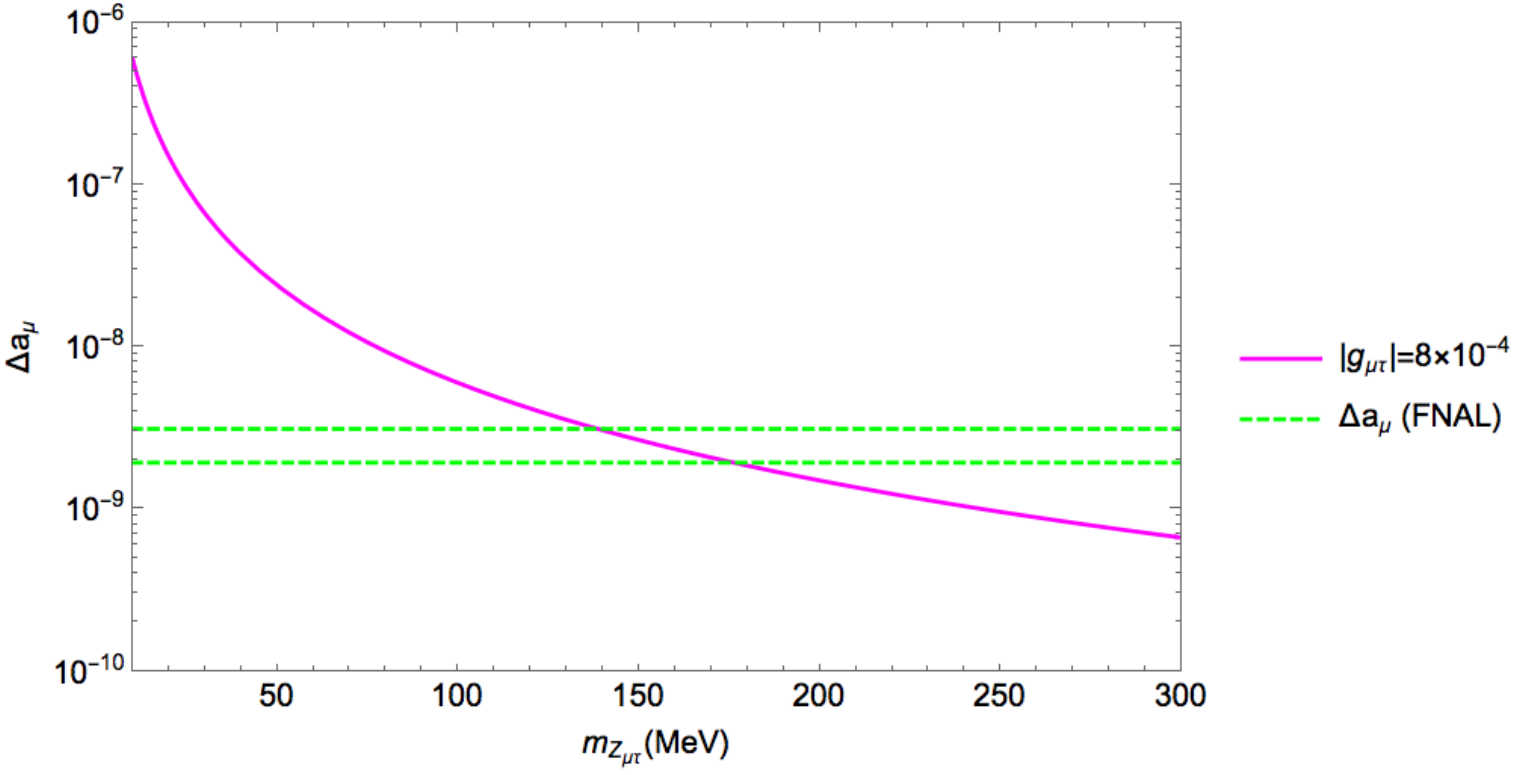}
\caption{Plot showing the contribution coming from new gauge boson $Z_{\mu \tau}$ $vs$ mass of $Z_{\mu \tau}$.}
\label{fig:numestZmt}
\end{figure}

(c) Contribution due to extra gauge boson, $Z_{\mu \tau}$ mediation :
\begin{equation}
\Delta a_\mu (Z_{\mu \tau}) = \frac{g^2_{\mu \tau}}{12\pi^2}\frac{m^2_\mu}{m^2_{Z_{\mu \tau}}},
\label{eqn:anaexpZmt}
\end{equation}
where $m_{Z_{\mu \tau}}$ and $g_{\mu \tau}$ are the mass of the $Z_{\mu \tau}$ and coupling 
between the $Z_{\mu \tau}$ and $\mu$, respectively.

The contributions coming from new gauge boson $Z_{\mu \tau}$ to $\Delta a_\mu$ is presented in 
figure \ref{fig:numestZmt}. Both $m_{Z_{\mu \tau}}$ and $g_{\mu \tau}$ are strongly constrained by 
the measurement of neutrino trident cross section by experiments 
like CHARM-II \cite{GEIREGAT1990271} and CCFR \cite{PhysRevLett.66.3117}.
The present limits are $m_{Z_{\mu \tau}}$ in the range $(100-200)$ MeV
and $g_{\mu \tau} \leq 10^{-3}$. In our numerical analysis we have fixed  
$g_{\mu \tau} = 8 \times 10^{-4}$ (just below the experimental limit) and varied
the $m_{Z_{\mu \tau}}$ in its allowed range. We see that $m_{Z_{\mu \tau}}$ in 
the range $(140-190)$ MeV can account for the entire anomaly. The $Z_{\mu\tau}$ exchange is the mechanism to explain the muon $(g-2)$ anomaly in our model because the contributions from both the heavy gauge boson exchange and the heavy scalar exchange are negligibly small.

\section{Charged Lepton Flavor Violation}
\label{sec:LFV}
Our model predicts observably large charged lepton flavor violation (cLFV) in the $\mu-\tau$ sector. Here we consider the radiative decay $\tau \rightarrow \mu \gamma$ whose branching ratio has the upper limit $\text{BR} (\tau \rightarrow \mu \gamma)$ is $< 1.5 \times 10^{-8}$ \cite{BaBar:2009hkt}. This decay can occur either through $W_L$ and $W_R$ mediation or through charged scalar mediation. The $U(1)_{L_\mu - L_\tau}$ symmetry forbids this decay through both $Z_R$ and $Z_{\mu\tau}$ mediation and through neutral scalar mediation. 

\begin{figure}[htb!]
    \centering
    \includegraphics[scale=0.6]{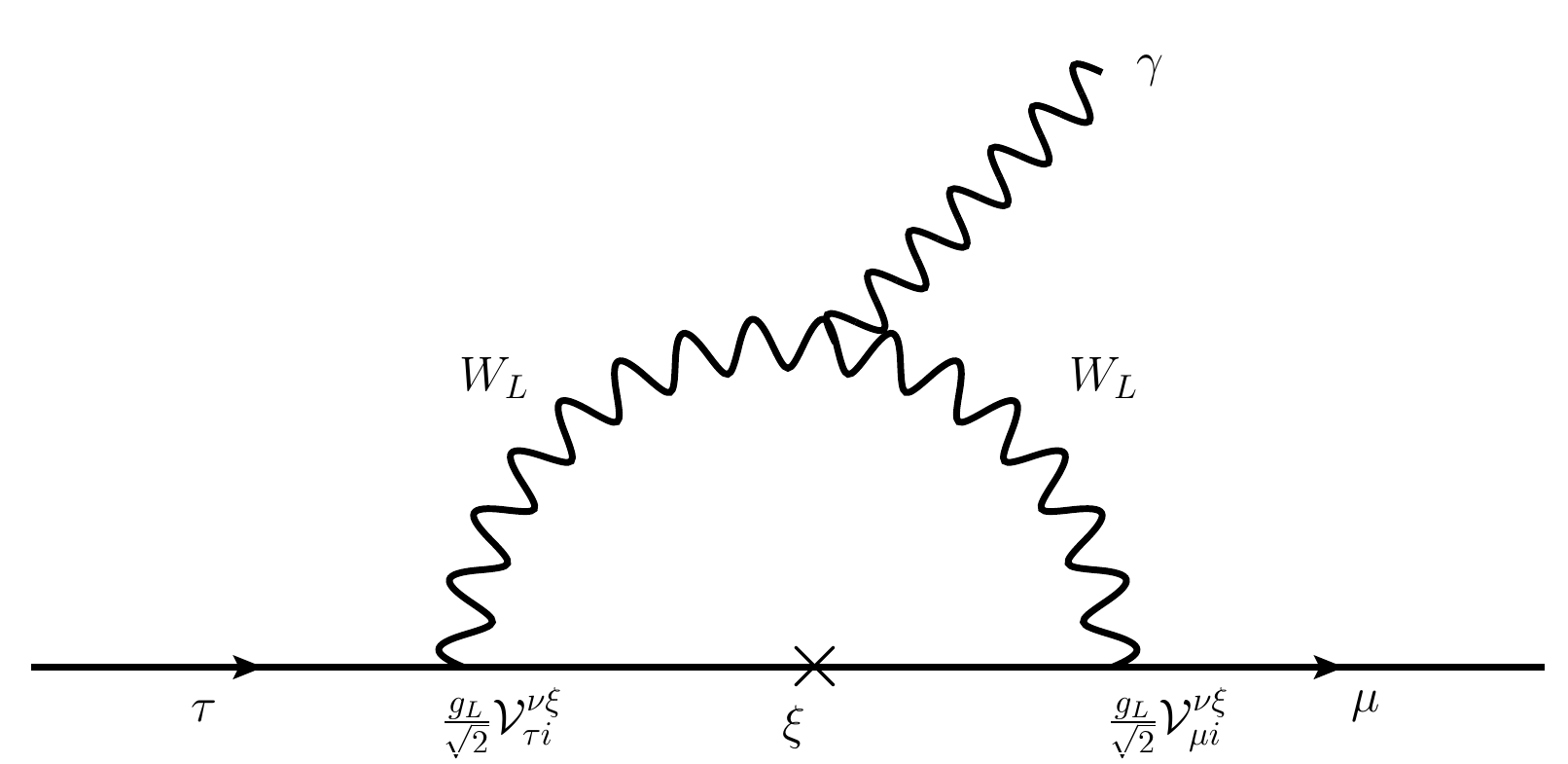}
    \caption{Feynman diagram for $\tau \rightarrow \mu \gamma$ process.}
    \label{fig:LFVdia}
\end{figure}

The Feynman diagram for $\tau \rightarrow \mu \gamma$ through $W_L$ mediation is shown in figure \ref{fig:LFVdia}. The expression for the branching ratio is given by \cite{Lindner:2016bgg}
\begin{equation}
    \text{BR} (\tau \rightarrow \mu \gamma) \sim \frac{3 (4\pi)^3 \alpha_{em}}{4G_F^2} \left[ \left| A^M_{\mu \tau}\right|^2 +  \left| A^E_{\mu \tau}\right|^2 \right],
    \label{eq:LFV}
\end{equation}
where $G_F$ is the Fermi's constant of weak interactions and $\alpha_{em}$ is the electromagnetic fine structure constant. The quantities $A^M_{\mu\tau}$ and $A^E_{\mu\tau}$ are the magnetic and electric dipole transition amplitudes.

In the $W_L$ mediated diagram we have $ m_\xi \simeq M_{W_L} \gg m_\tau$. In this limit, we have 
\begin{align}
    &  A^M_{\mu \tau} = i A^E_{\mu \tau} \simeq - \frac{1}{16\pi^2 M_{W_L}^2} \left(\frac{g_L^2}{2}\right) \left(\frac{17}{12}\right) \mathcal{V}_{\tau i}^{\nu \xi} \mathcal{V}_{\mu i}^{\nu \xi} \\
    & \implies \text{BR} (\tau \rightarrow \mu \gamma) \simeq 6.43 \times 10^{-6} \left(\frac{1~\rm{TeV}}{M_{W_L}}\right)^4 \left| \mathcal{V}_{\tau i}^{\nu \xi} \mathcal{V}_{\mu i}^{\nu \xi} \right|^2 \simeq 
    0.16 \left| \mathcal{V}_{\tau i}^{\nu \xi} \mathcal{V}_{\mu i}^{\nu \xi} \right|^2.
    \label{eq:WLcLFV}
\end{align}
The predicted branching ratio is of the order of the present upper limit if 
\begin{equation}
   \left| \mathcal{V}_{\tau i}^{\nu \xi} \mathcal{V}_{\mu i}^{\nu \xi} \right|^2 < 10^{-7} .
   \label{eq:upperlimit}
\end{equation}
 In the previous section we noted that the experimental bound on the deviation from unitarity in muon sector is $(|V_{\mu i}^{\nu \xi}|^2) \leq 1.6 \times 10^{-3}$. The corresponding limit in the tau sector is $(|V_{\tau i}^{\nu \xi}|^2) \leq 5.4 \times 10^{-3}$ \cite{Antusch:2008tz,Antusch:2009pm,Antusch:2006vwa,Forero:2011pc}. The values of light-heavy neutrino mixing needed to obtain observably large branching ratio of $\tau \rightarrow \mu \gamma$, given in Eq. \ref{eq:upperlimit}, are well within these non-unitarity bounds. 

For $W_R$ mediated diagram we will have expressions for $A^{M}_{\mu\tau}$, $A^{E}_{\mu\tau}$ and $\text{BR} (\tau \rightarrow \mu \gamma)$ similar to those in Eq. \ref{eq:WLcLFV}, with $M_{W_L}$ replaced by $M_{W_R}$. Since $M_{W_R}$ is at least an order of magnitude greater than $M_{W_L}$, this contribution is expected to be very small. 

\section{Estimation of neutrino masses}
\label{sec:numass}
In this section, we consider the constraints our model imposes on the parameters of light neutrino sector. In the framework of the LISS mechanism described in section \ref{sec:mod}, the mass matrix for the neutral fermions $M_D$, given in Eq. \ref{eq:MD}, and the $N_R-S_L$ coupling matrix $M$, given in Eq. \ref{eqn:LNS}, are diagonal. The explicit form of light neutrino mass matrix $m_\nu$ is given in Eq. \ref{eq:mnu}. This matrix is non-diagonal because the matrix $\mu_S$ is non-diagonal. 

As discussed in section \ref{sec:mod} the $U(1)_{L_\mu-L_\tau}$ symmetry of the model requires 
\begin{equation}
(\mu_S)_{\mu \mu} = 0 = (\mu_S)_{\tau \tau} ~ \implies ~ (m_\nu)_{\mu \mu} = 0 = (m_\nu)_{\tau \tau}.
\end{equation}
That is : the effective masses of $\nu_\mu$ and $\nu_\tau$ should vanish in this model. The 
different textures of neutrino mass matrices, with two zero elements are classified in~\cite{Frampton:2002yf}. The texture with $(m_\nu)_{22}) = 0 = (m_\nu)_{33})$ is labelled
"C" in that classification. The compatibility of different two zero textures with precision 
neutrino oscillation data is studied in~\cite{Alcaide:2018vni}. The four solutions, shown in table~\ref{tab:textureC}, give the predictions of their fit for the smallest neutrino mass,
sum of light neutrino masses, the effective mass for neutrinoless double beta decay and the phases
$\delta$, $\alpha_1$ and $\alpha_2$.

\begin{table}[htb]
\begin{center}
\begin{tabular}{|c|c|c|c|c|c|c|}
 \hline
Hierarchy & $m_{\rm{min}}$ & $\sum_i m_i$ & $m_{\rm{ee}}$ & $\delta$ & $\alpha_2$ & $\alpha_3$  \\
\hline
 NH & $> 159$ & $> 484$ & $> 151$ & $175-262$ & $178-180$ & $178-180$ \\
 \hline
 NH & $> 167$ & $> 484$ & $> 151$ & $278-346$ & $180-182$ & $180-182$ \\
 \hline
 IH & $> 35$ & $> 155$ & $> 34$ & $231-269$ & $178-180$ & $151-178$ \\
 \hline
 IH & $> 67$ & $> 155$ & $> 34$ & $273-289$ & $120-168$ & $185-202$ \\
 \hline
 \end{tabular}
\end{center}
\caption{Predictions of~\cite{Alcaide:2018vni} for the two zero texture "C". All the masses
are given in units of milli-eV and all the phases are given in degrees.}
\label{tab:textureC}
\end{table}

We note that in all four cases the sum of light neutrino masses exceeds the 
cosmological upper bound of $0.11$ eV~\cite{Planck:2018vyg}. The violation of the upper bound is
much more modest in the case of IH than in the case of NH. The symmetries of the model impose a particular
two-zero texture on the light neutrino mass matrix. The texture constraints, when combined with neutrino oscillation data and the cosmological bound on the sum of light neutrino masses, show a strong preference for inverted hierarchy. They
also require $\delta$, the CP violating phase in neutrino 
oscillations, to be close to maximal.

The neutrino data
plus the constraints of the model strongly prefer inverted hierarchy. They also
predict a value of $\delta$ which leads to a large CP violation in neutrino 
oscillations.

\section{Conclusion}
\label{sec:conclusion}
We have studied an $U(1)_{L_\mu -L_\tau}$ extended left-right theory. In this framework the neutrino masses are generated via inverse seesaw scenario. We did a numerical analysis of muon $(g-2)$ anomaly in this model. Given the stringent constraints on deviation from unitarity on the mixings of active lepton flavors, we found that the muon $(g-2)$ anomaly \textit{cannot} be explained through the mediation of either heavy gauge bosons or heavy scalars. However the model contains a neutral light gauge boson $Z_{\mu\tau}$ arising from the gauged $U(1)_{L_\mu-L_\tau}$ symmetry. Both the mass and couplings of this gauge boson are strongly constrained by the neutrino trident cross section. The muon $(g-2)$ anomaly can fully be accounted for the gauge coupling strength $g_{\mu\tau} = 8 \times 10^{-4}$ and $m_{Z_{\mu\tau}}$ in the range $(140-190)$ MeV, both of which are allowed by present data. 

The model also predicts observable cLFV in $\mu-\tau$ sector. We studied the decay $\tau \rightarrow \mu \gamma$ in this model and found that an observable branching ratio is possible even while satisfying the stringent constraints on light-heavy neutrino mixing. Other interesting charged lepton flavor violation will be considered in future work.

The $U(1)_{L_\mu - L_\tau}$ symmetry of the model imposes the following constraints on the light neutrino mass matrix : $(m_\nu)_{\mu \mu} = 0 = (m_\nu)_{\tau \tau}$. These two constraints can be satisfied simultaneously only for some very specific values of light neutrino masses. In the case of NH, the allowed value of the lightest neutrino mass is $m_1 = 0.16$ eV which leads to a strong 
violation of the cosmological upper bound on the sum of light neutrino masses. In the case of IH, the allowed value of the lightest neutrino mass are $m_3 =  0.035$ eV, which leads to a much milder violation of the cosmological upper bound. In the latter case, the combination of model constraints
and the neutrino data require the value of $\delta$ to be in the range which leads to a large CP violation in neutrino oscillations. The Majorana phases $\alpha_1$ and $\alpha_2$ are constrained to be close to $180$
so that the necessary cancellations in $(m_\nu)_{\mu \mu}$ and $(m_\nu)_{\tau \tau}$ can take place.
Thus the model predicts moderate values for the minimum neutrino mass and the effective mass
$(m_\nu)_{ee}$ for neutrinoless double beta decay but predicts a very large CP violation in neutrino oscillations.

\section{Acknowledgement}
CM and SS acknowledge Institute Postdoctoral Fellowship of IIT Bombay for financial support.

\bibliographystyle{utcaps_mod}
\bibliography{muon}
\end{document}